# Protein domains as units of genetic transfer

Cheong Xin Chan[1], Robert G. Beiko[2], Aaron E. Darling[1] and Mark A. Ragan[1]

[1]ARC Centre of Excellence in Bioinformatics and Institute for Molecular Bioscience, The University of Queensland, Brisbane, QLD 4072, Australia

[2]Faculty of Computer Science, Dalhousie University, 6050 University Avenue, Halifax, Nova Scotia, Canada B3H 1W5

In prokaryotes, breakpoints of genetic recombination significantly demarcate genomic regions that encode protein domains.

**Abstract**

Genomes evolve as modules. In prokaryotes (and some eukaryotes), genetic material can be transferred between species and integrated into the genome via homologous or illegitimate recombination. There is little reason to imagine that the units of transfer correspond to entire genes; however, such units have not been rigorously characterized. We examined fragmentary genetic transfer in single-copy gene families from 144 prokaryotic genomes and found that breakpoints are located significantly closer to the boundaries of genomic regions that encode annotated structural domains of proteins than expected by chance, particularly when recombining sequences are more divergent. This correlation results from recombination events themselves and not from differential nucleotide substitution. We report the first systematic study relating genetic recombination to structural features at the protein level.

**Introduction**

Genomes are shaped by processes that direct the descent of genetic material. The main process has been considered to be vertical (parent-to-offspring) descent within a genomic lineage. More recently, the role of lateral genetic transfer (LGT) has been emphasized, particularly among the prokaryotes [1-3], in contributing to the origin of physiological diversity [4]. A transfer event involves the acquisition of external genetic fragments into the cell and their subsequent integration into the host chromosome through recombination. These recombined regions might correspond to complete genes, multi-gene clusters [5], or fragments of genes [6]. Breakpoints might thus be located in a random pattern along the genome, or be positively or negatively associated with boundaries of regions that encode structural units.

The genomes of prokaryotes have small intergenic regions and consist largely of protein-coding sequences. The proteins so encoded often consist of one or more spatially compact structural units known as *domains* which may fold autonomously and, singly or in combination, convey the protein's specific functions [7, 8]. As natural selection is based on function, we examined the possibility that domains also serve as units of genetic transfer, *i.e.* whether the transferred regions correspond to the intact structural domains of proteins.

We showed earlier, by phylogenetic analysis of 22437 putatively orthologous protein families of 144 fully-sequenced prokaryotic genomes [9], that vertical transmission is the dominant mode of genetic inheritance in prokaryotes, but LGT has contributed significantly to the composition of some genomes. Comparison between phylogeny inferred for each protein family and a reference organismal phylogeny implied, at a posterior probability threshold of 95% or greater, that about 13.4% of the tested relationships (bipartitions) have been affected by LGT. In that study, we treated each protein (gene) as a unit. The dataset developed for that study provides a unique platform to test whether transferred genetic regions in prokaryotes correspond to intact structural domains of proteins. Our null hypothesis is that no such correlation exists.

**Units of genetic transfer**

We implemented a two-phase strategy [10] for the detection of recombination events among these data. To remove potential complication from paralogous history in the sequences and to ensure a confident inference of genetic transfer event rather subsequent evolution of duplicated genes, we selected 1462 single-copy gene families for which no gene is duplicated within the corresponding genome; family sizes ranged from 4 to 52. We first applied three statistical methods [11] to detect recombination events, then identified recombination breakpoints via a rigorous Bayesian phylogenetic approach [12] that infers changes in tree topologies and evolutionary rates across sites within a sequence set. The Bayesian approach has been shown to perform at high accuracy in delineating breakpoints [13]. On this basis we classified the gene families into five categories based on support for alternative topologies and width (number of alignment positions) of the transition between topologies (Table 1). Sequence sets presenting clear evidence of recombination within the gene boundaries were categorized into Classes A (1.6%), B (9.3%) and C (8.6%), with Class A showing abrupt changes in Bayesian posterior probability (BPP) support for alternative topologies in the breakpoint region indicative of recent transfer, Class B showing a more gradual change in such BPP indicative of a less-recent transfer or incomplete taxon sampling, and Class C exhibiting both abrupt and gradual changes in BPP. Sequence sets with inconclusive evidence were grouped as Class D (5.5%), and those with no evidence of recombination as Class E (75%).

Table 1. Classification of results in breakpoint identification. The criteria used in the classification are support (Bayesian posterior probability, BPP) for alternative tree topologies in the breakpoint region, and number of aligned nucleotide positions (nt) over which the topology changes. Cases in which all breakpoints show abrupt change between very strongly supported topologies constitute Class A, and those in which all breakpoints show more-gradual change between moderately to strongly supported topologies constitute Class B. Class C groups individual cases showing both abrupt and more-gradual BPP changes across breakpoints. Classes A-C represent positively identified recombination events, and precise breakpoints were inferred. Cases showing inconclusive support (BPP <0.50) at breakpoint regions were classified as Class D, and those that show no change were classified as Class E.

| Classes | A | B | C | D | E |
| --- | --- | --- | --- | --- | --- |
| Support (BPP) of alternative tree topologies in breakpoint region | ≥ 0.90 | ≥ 0.50 | ≥ 0.50 | < 0.50 | None |
| Region length (nt) over which BPP change occurs | ≤ 30 | > 30 | Mixed | Any | None |
| Inference of recombination | + | + | + | ? | - |

To investigate the correlation of recombination breakpoints with boundaries of protein structural domains, we applied a breakpoint distance-to-boundary statistic adapted from previous studies [14, 15] with distance assessed as the number of aligned amino acid positions between a breakpoint and the nearest domain boundary. Domains, in this study, are defined as evolutionary units of proteins as inferred by structural and/or sequence homology in the protein databases of Pfam [16] and SCOP [17]. The results for Classes A, B and C are shown in Fig. 1. We found that 16.0% (Class A), 40.6% (B) and 30.4% (C) of identified recombination breakpoints are located within 20 amino acids of protein structural domain boundaries, and for all these classes > 50% fall within 50 amino acid positions (Fig. 1).

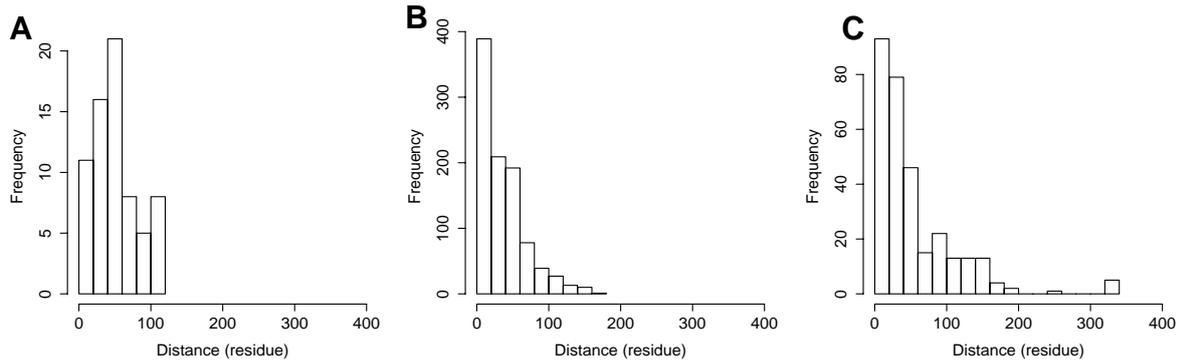

Fig. 1. Distances between inferred breakpoint and the nearest protein domain boundary for (A) Class A, (B) Class B and (C) Class C. For each graph, the X-axis represents the number of amino acids between the breakpoint and the nearest domain boundary, and the Y-axis represents the number of breakpoints. Protein domain boundaries were determined by homology search against domain entries in Pfam [16] and SCOP [17]. Distance distributions in each of A, B and C were compared against a distribution of breakpoint-boundary distances averaged from 100,000 randomly permuted breakpoints within the same dataset using the Kolmogorov-Smirnov test [18].

For each of Class A, B and C, the distribution of observed breakpoint-boundary distances (Fig. 1) was compared to a distribution of expected breakpoint-boundary distances averaged from 100,000 randomly permuted breakpoints within the same dataset. Each permutation was carried out by assigning a random recombination breakpoint in the sequence and calculating the distance of the breakpoint to the closest domain boundary that is annotated on the sequence. Using the one-sided Kolmogorov-Smirnov test [18], we rejected the expected distribution (null hypothesis) for each class at a significance level of $\leq 10^{-13}$, strongly implying that observed distances between breakpoints and boundaries of protein domains are significantly shorter than expected by chance. Fig. 2 shows a quantile-quantile plot between these observed distances and those expected from permutation for each class.

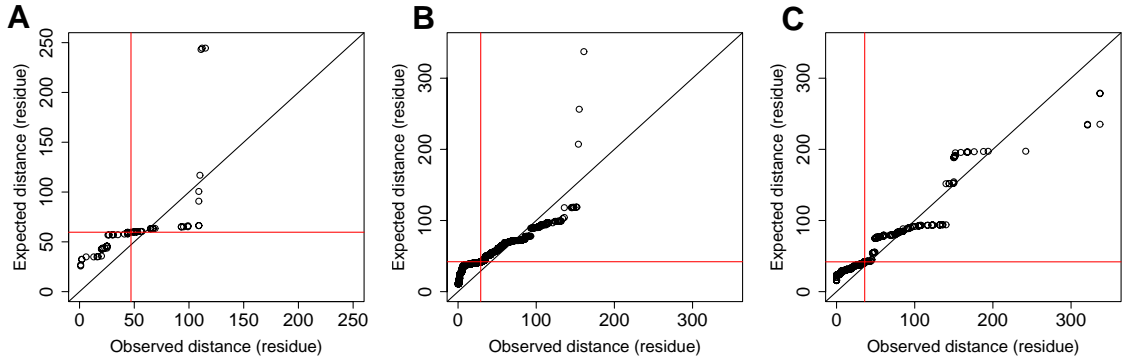

Fig. 2. Quantile-quantile plots of observed and expected breakpoint-boundary distances distances for (A) Class A, (B) Class B and (C) Class C. For each graph, the X- and Y-axes represent the observed and expected breakpoint-boundary distances respectively, counted by number of amino acid residues. The red vertical and horizontal lines show quantile 0.5 for each axis. If the distribution of observed distances is not different from the distribution of expected distances, the distances will have a one-to-one relationship in the quantile-quantile plot (as shown by the diagonal line passing through the origin). At least half (quantile 0.5) of the observed breakpoint-boundary distances are smaller than the expected distances for all classes. Complementing the Kolmogorov-Smirnov tests, the results indicate that the observed breakpoints are located closer to protein domain boundaries than one would expect by chance.

At least half of observed breakpoint-boundary distances (at quantile 0.5) are significantly different and smaller than expected, supporting the observation in Fig. 1. For each identified breakpoint in each class, we measured the proportion of the observed distance of the breakpoint from the nearest domain boundary as a fraction of the length of the region at which that breakpoint can possibly be defined, and termed this measure as $\rho$, where $\rho = [0,1]$, as shown in Fig. 3. If the breakpoints are located at random, values of $\rho$ are expected to be normally distributed around the mean 0.50. For each class A, B and C, we observed the mean of $\rho$ to be less than 0.50 (Class A: $0.39 \pm 0.055$; Class B: $0.35 \pm 0.016$ and Class C: $0.27 \pm 0.022$), where the $\pm$ values give the 95% confidence interval estimated from a $t$-distribution. These results therefore show that there is a significant bias toward recombination breakpoints being located near boundaries of protein domains for all classes, consistent with domains being largely preserved during (and after) genetic transfer. The deviation from expectation was stronger in the less-recent fragmentary transfer events (Class B) and in cases most readily interpreted as an overlay of recent and less-recent transfer (Class C), compared to recent transfer events (Class A). However, this may be a consequence of the smaller sample size in Class A (69) compared to Classes B (958) and C (306).

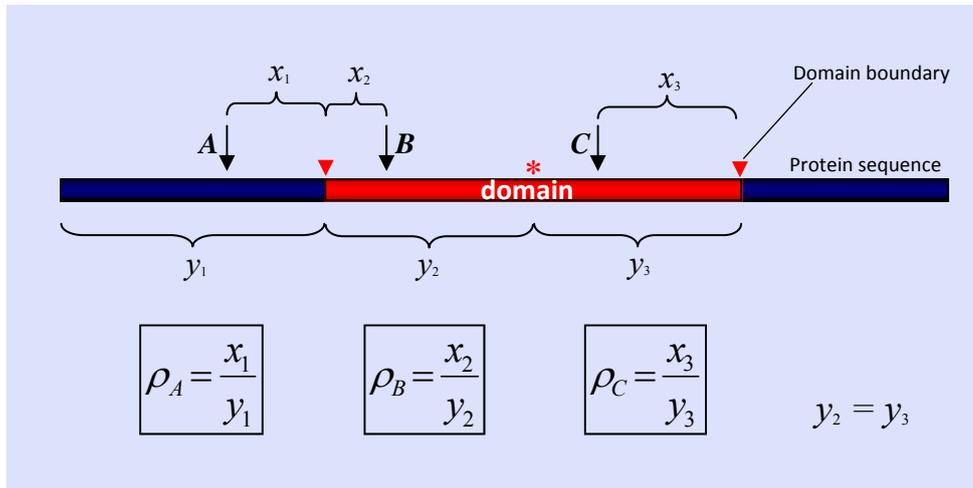

Fig. 3. Calculation of $\rho$. A protein sequence is illustrated with a single domain (red region) in the middle. The two domain boundaries are represented by the red triangles. Three breakpoints are illustrated with the black arrows on the sequence: $A$, $B$ and $C$. A $\rho$ value represents the proportion of the observed distance of a breakpoint from the nearest domain boundary ($x$) in respect to the length of the region within which that breakpoint can possibly be defined ($y$). The respective calculation of $\rho$ for breakpoints $A$, $B$ and $C$ are shown. The $\rho$ values range between 0 and 1. The red asterisk represents the midpoint of the domain, therefore $y_2 = y_3$.

To examine the effect of sequence divergence on the breakpoint location in relation to domain boundaries, we re-classified the gene families for which breakpoints are inferred into three categories based on the divergence of recombining sequences in the families. Fig. 4 shows the observed breakpoint-boundary distances based on the divergence of recombining sequences in the dataset: low divergence (sequence similarity $\geq 0.50$); moderate divergence (sequence similarity between 0.3 and 0.5); and high divergence (sequence similarity $< 0.3$). For each category of divergence, the distribution of the observed breakpoint-boundary distances was compared against the distribution of expected breakpoint-boundary distances that was averaged from 100,000 randomly permuted breakpoints within the same dataset, and the quantile-quantile plot for each comparison is shown. All three observed distributions are different from the expected distributions at a significance level of $< 10^{-16}$.

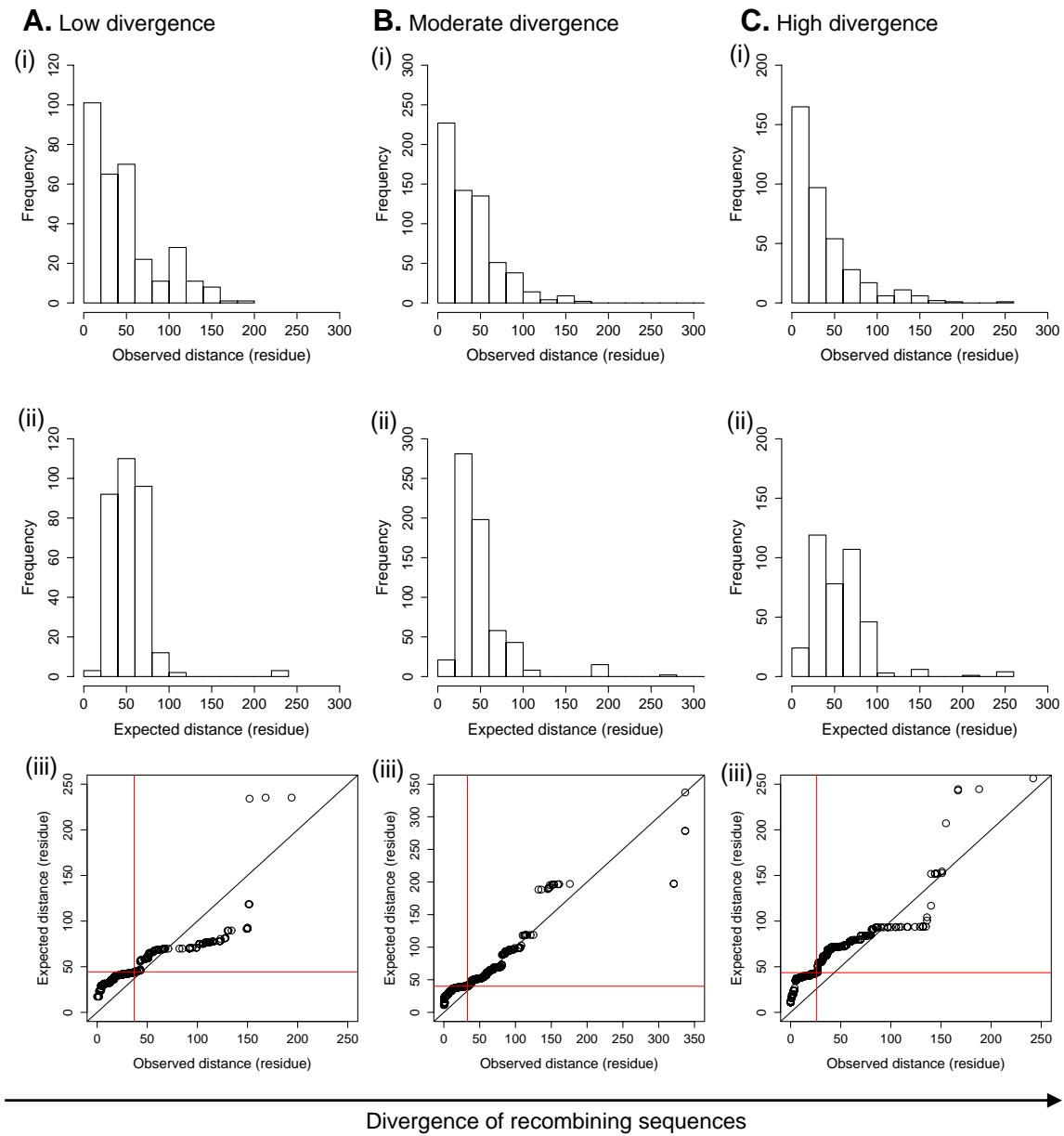

Fig. 4. Distances between inferred breakpoint and the nearest protein domain boundary based on divergence of recombining sequences: (A) low, (B) moderate and (C) high divergence. The panels of (i) and (ii) show the observed distances and expected distances, respectively. For each graph in (i) and (ii), the X-axis represents the number of amino acids between the breakpoint and the nearest domain boundary, and the Y-axis represents the number of breakpoints. The expected distances in (ii) were averaged from 100,000 randomly permuted breakpoints. The panels of (iii) show the quantile-quantile plots of observed and expected breakpoint-boundary distances distances for each category of divergence. For each graph, the X- and Y-axes represent the observed and expected breakpoint-boundary distances respectively, counted by number of amino acid residues. Red vertical and horizontal lines show quantile 0.5 for each axis. If the distribution of observed distances is not different from the distribution of expected distances, the distances will have a one-to-one relationship in the quantile-quantile plot (as shown by the diagonal line passing through the origin).

As shown in the quantile-quantile plots (panels of (iii) in Fig. 4) across all three categories, at least half of the observed breakpoint-boundary distances (quantile 0.5) are shorter than expected, particularly in cases when the divergence of the recombining sequences is high. Thus when the recombining sequences are more divergent, we observe a greater number of breakpoints located closer to domain boundaries than expected.

**Breakpoint location is the result of recombination**

Our observations raise the question of whether the bias of breakpoint location results from selection against nucleotide substitution within the transferred region that encodes the domain, with accumulation of substitutions outside, or alternatively arises from the recombination event *per se*. The cases of recent transfer into the gene (Class A) cannot be explained by selection: the abrupt change in BPP support for alternative topologies in the breakpoint region indicates a recent event leaving insufficient time for substitutions to accumulate within domain-flanking regions. For cases of less-recent transfer (Classes B and C), invoking selection on substitution processes to explain the more-gradual change of BPP would imply substantially different substitution rates between introgressed and background sequences; this is not the case, as using a Bayesian approach [12] we found no instance in which substitution rates differed by more than 0.30 across the entire alignment. Nor can the proximity of breakpoints to domain boundaries be attributed to the truncation (or extension) of domains, as the mean length of homologous domains is not significantly different (*p*-value = 0.48) in the presence of recombination (Classes A-C) compared to its absence (Class E).

**Concluding remarks**

Genomes evolve in modular fashion, with different evolutionary histories for different regions [1, 3]. Our work shows that LGT can produce genes with mosaic ancestries and that the units of such transfer are significantly correlated to protein domains. Other LGT may transfer entire genes or groups of genes, although these cases are not detected by the methods we applied here. The relationship between structural domains and units of fragmentary gene transfer suggests that not only genomes, but also the function of individual proteins, can evolve in a modular fashion through LGT.

**Acknowledgements**

This study is supported by an Australian Research Council grant CE0348221. We thank Vladimir Minin for his inputs on DualBrothers, Lynn Fink for technical assistance and Lloyd Flack for useful discussions on statistics. CXC is supported by a UQIPRS scholarship; AED is supported by NSF grant DBI-0630765.